\documentclass[12pt,preprint]{aastex}

\slugcomment{To appear in 2003, ApJ Letters, 586, L145}

\shorttitle{X-ray emission morphology and rotation in solar-type stars}
\shortauthors{Barnes}

\begin{document}


\title{A connection between the morphology of the X-ray emission\\ 
and rotation for solar-type stars in open clusters\footnote{This is paper 14 
in the WIYN Open Cluster Study (WOCS).}
}


\author{Sydney A. Barnes}
\affil{Astronomy Department, Yale University, New Haven, CT 06520, USA}
\email{barnes@astro.yale.edu}




\begin{abstract}

It is suggested that the three-segmented morphology of the soft X-ray emission 
from cluster and field stars may be understood in terms of the recent
classification of rotating stars into three kinds: those lying on the 
{\it convective} sequence, on the {\it interface} sequence, or in the 
{\it gap} between them.

\end{abstract}


\keywords{open clusters and associations: general --- stars: activity ---
 stars: chromospheres --- stars: late-type --- stars: magnetic fields --- 
 stars: rotation}



\section{Introduction} \label{intro}

The soft ($ < 2.5 $ keV) X-ray emission from late-type cluster stars is 
currently believed to delineate three kinds of dependence on stellar rotation 
or more precisely, Rossby number. 
This was originally noted in Patten \& Simon (1996), was reviewed in Randich 
(2000), and has been reiterated recently by Feigelson et al. (2003).
Other studies have noted similar results for cluster and field stars 
(eg. James et al. 2000).
In the regime of slow rotation, a clear rotation-activity relationship is
observed, with X-ray emission declining steadily with increasing rotation
period. 
For faster rotators, there is evidence of X-ray ``saturation,'' a phenomenon
describing a relative constancy of the X-ray emission at an elevated level
regardless of increasing rotation speed.
Finally, for the very fastest rotators, there appears to be a moderate
decline in the X-ray emission, and this phenomenon is sometimes called
``super-saturation.'' 
The physical basis for these effects is presently unclear.

It has recently been proposed that the morphology of the
rotational observations of cluster and field stars may be understood in terms
of a magnetic paradigm (Barnes 2003). 
According to this paradigm, the fastest rotators in open clusters lie on a 
{\it convective} sequence, where a convective magnetic field dominates, and 
the outer convection zone of the stars shears past the interior radiative zone.
It is proposed that this shear generates an interface dynamo which attempts
to couple the two zones, driving the star through a {\it gap} (in rotation) 
region, towards the {\it interface} sequence, one occupied by stars with 
well-developed dynamos, and including the present-day Sun. 

If this paradigm is valid, it should also be reflected in the X-ray
properties of stars, since these are believed to originate in stellar
magnetism. 
Here, I attempt to show that in the case of specific open clusters
where high-quality rotation periods and X-ray data are both available,
the X-ray emission does, in fact, conform to the new paradigm.


\section{Observations} \label{obs}

A meaningful test of the paradigm requires clusters that have not only been
well-studied in X-rays but where the cluster stars can be placed on 
the underlying rotational sequences with absolute unambiguity.
This latter requirement implies the use of rotation periods.

Four such groups of stars are included here - 
{\sc IC\,2391: } Patten \& Simon (1996);
{\sc IC\,2602: } Barnes et al. (1999);
{\sc Alpha Per: } Prosser \& Grankin (1997);
and a group of field M dwarfs from James et al. (2000).
Using the $B-V$ color and rotation period, it is possible to classify each 
star in this set in terms of the paradigm advanced in Barnes (2003) as lying
on the convective sequence (C), interface sequence (I), or in the rotational 
gap (g) between these sequences. 
This classification is displayed graphically in the left-hand panels of 
Figure\,1. 
It is based on the rotation period of the star in question, taking account of 
its (dereddened) color and age, except in the case of the M\,dwarfs whose 
heterogeneous ages do not permit an unambiguous identification with the 
relevant sequence. An ambiguity discussed thoroughly 
in Barnes (2003) with respect to very late-type stars suggests a more cautious
labelling of the three slowest rotators in this set merely by the label ``S.''

Having classified each star as C, g, or I/S, we consider the X-ray properties.
For each of these stars, the soft X-ray emission values are available from
the following studies - 
{\sc IC\,2391: } Patten \& Simon (1996);
{\sc IC\,2602: } Randich et al. (1995);
{\sc Alpha Per: } Randich et al. (1996) and Prosser et al. (1996); 
and for field M\,stars:  James et al. (2000).
In the right-hand panels of Figure\,1, we correspondingly plot the X-ray 
emission from each of these stars\footnote{James et al. (2000) suggest that
some caution is warranted in interpreting data where the bolometric corrections
for late-type stars are calculated using $B-V$ colors.}
against the Rossby Number, $N_R = P/\tau_c$, where $P$ is the rotation period 
of the star and $\tau_c$ is the convective turnover timescale, estimated using 
the relationship from Kim and Demarque (1996). 
In these plots, as before, we mark each star by the symbol C, g, or I/S, 
based on the classification described earlier.

One observes that the stars marked C lie in the super-saturated region
of X-ray activity, those marked g in the saturated region, and those 
marked I in the region of validity of the rotation-activity relationship.

This picture is further reinforced by Figure\,1 in Randich (2000), which shows
that all the solar-type field stars, expected to lie on the Interface sequence
(I) in the new paradigm, obey the standard rotation-activity relationship.


\section{Interpretation}

The foregoing observations encourage us to think that the paradigm suggested 
to account for the rotational evolution of stars might also be useful in 
understanding the X-ray emission from stars.

Purely observationally, we know that the X-ray emission  
$ X = X(N_R) = X(P/\tau_c)$.
Now we know from Barnes (2003) that $P$ is not a smoothly varying function
of stellar properties but rather that
\begin{equation}
P = \cases
	{P_C	& for stars on the convective sequence;\cr
  	 P_g	& for stars in the gap;\cr
	 P_I	& for stars on the interface sequence.\cr }
\end{equation}
where $P_C$ and $P_I$ have separate functional dependences on stellar
age and color, and $P_g$ merely represents stars with rotation periods in 
the gap between the Convective and Interface sequences. 
Very briefly, $P_I \sim \sqrt{t/I_s}$ but $P_C \sim e^{t/(KI_c)}$ where
$t$ is the age, $I_s$ and $I_c$ are, respectively, the moments of inertia 
of the entire star or only the convection zone, and $K$ is a constant.
The origin of these sequences in terms
of stellar properties has been explained at length in Barnes (2003).

Thus, the fact that the X-ray emission has the particular three-segmented
morphology that has been noted by various studies can be traced to
the fact that, for the three sections, we have 
\begin{equation}
X = \cases
	{X(P_C/\tau_c)	& for stars on the convective sequence;\cr
  	 X(P_g/\tau_c)	& for stars in the gap;\cr
	 X(P_I/\tau_c)	& for stars on the interface sequence.\cr}
\end{equation}

Although this scheme suggests the immediate reason for the differing
dependences on Rossby number, the underlying reason must have to do with
the differing magnetic fields on these sequences that are manifested in 
differing X-ray emission. 

If the X-ray emission scales simply with the total closed topology 
magnetic field, then we have a scheme here which roughly parallels the
suggestion in Barnes (2003) that $C \rightarrow g \rightarrow I$ represents
an evolutionary progression for solar-type stars.
In that case, the right-hand panels of Figure\,1 imply, in order,
the creation/strengthening of a magnetic field on the {\it C} sequence 
					(super-saturation in X-ray parlance), 
the attainment of a maximal field in the gap g 
					(saturation),
and the steady decay of this field for stars on the {\it I} sequence 
					(rotation-activity paradigm).

The paradigm suggested for the rotational evolution of stars is strikingly 
parallel. It was suggested that stars on the convective sequence are initially 
equipped with only a convective or small-scale magnetic field which cannot 
couple to the radiative interior. Thus, only the external convection zone is
braked. This results in a large shear at the convective/radiative interface,
which creates an interface-type dynamo and magnetic field. The field itself 
then couples the convective and radiative zones of the star, and drives the 
star through the gap to the interface sequence, where the entire star is 
braked subsequently. In this picture, the X-ray observations would therefore 
be interpreted to indicate the creation and strengthening of the interface
dynamo among the {\it C} sequence stars, 
the attainment of a maximal field for the {\it g} stars,
and the decay of the interface field among the {\it I} sequence stars.


\section{Chromospheric Ca\,II H\,\&\,K emission}

The Chromospheric Ca\,II H\,\&\,K emission from solar-type stars in the field
has been known to decline steadily with age (eg. Wilson 1963; Skumanich 1972).
In fact, this decline is well-behaved enough that it is currently used to 
derive stellar ages for field stars (Soderblom, Duncan \& Johnson 1991; 
Donahue 1998). These stars might reasonably be expected to be similar to
those that obey the rotation-Xray activity paradigm, and to be located on the
interface sequence since they are mostly older than several hundred megayears. 

In young clusters, however, there appear to be complex patterns of 
Ca\,II H\,\&\,K emission that cannot easily be mapped onto the standard 
age-rotation-activity paradigm (C.M. Anderson, 2002, personal communication).
If the new rotational paradigm is correct, it ought to be able to explain
these variations in a manner similar to the X-ray variations.
This task is currently underway.


\section{Conclusion}

The X-ray observations of cluster and field stars have a three-segmented
morphology largely because there are three kinds of rotating stars - 
those with only a convective magnetic field, here identified with the
super-saturation segment, 
those with a well-developed interface magnetic field, here identified with
those stars obeying the rotation-activity relationship, and
those in the process of conversion from the former to the latter state,
identified with the saturated activity segment.

Because rotating stars evolve systematically in the direction:
$ C \rightarrow g \rightarrow I $, the X-ray morphology of cluster and field
stars might merely reflect the build-up of large scale magnetic fields on
the convective sequence during this high-shear era, the attainment of a maximal
field (saturation) for the gap stars, and the gradual decline of X-rays with
magnetic field and rotation on the interface sequence.



\acknowledgments

SAB would like to acknowledge support from the NSF through AST-9986962.






\clearpage 

\begin{figure}
\plottwo{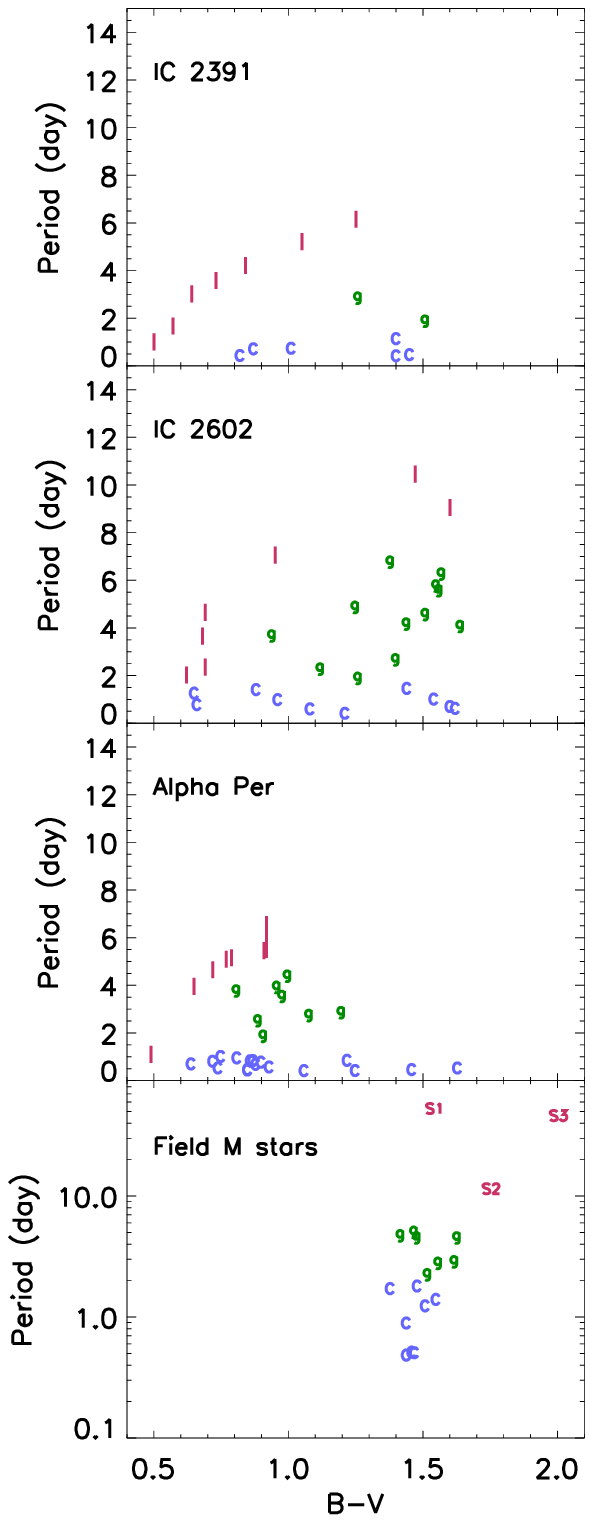}{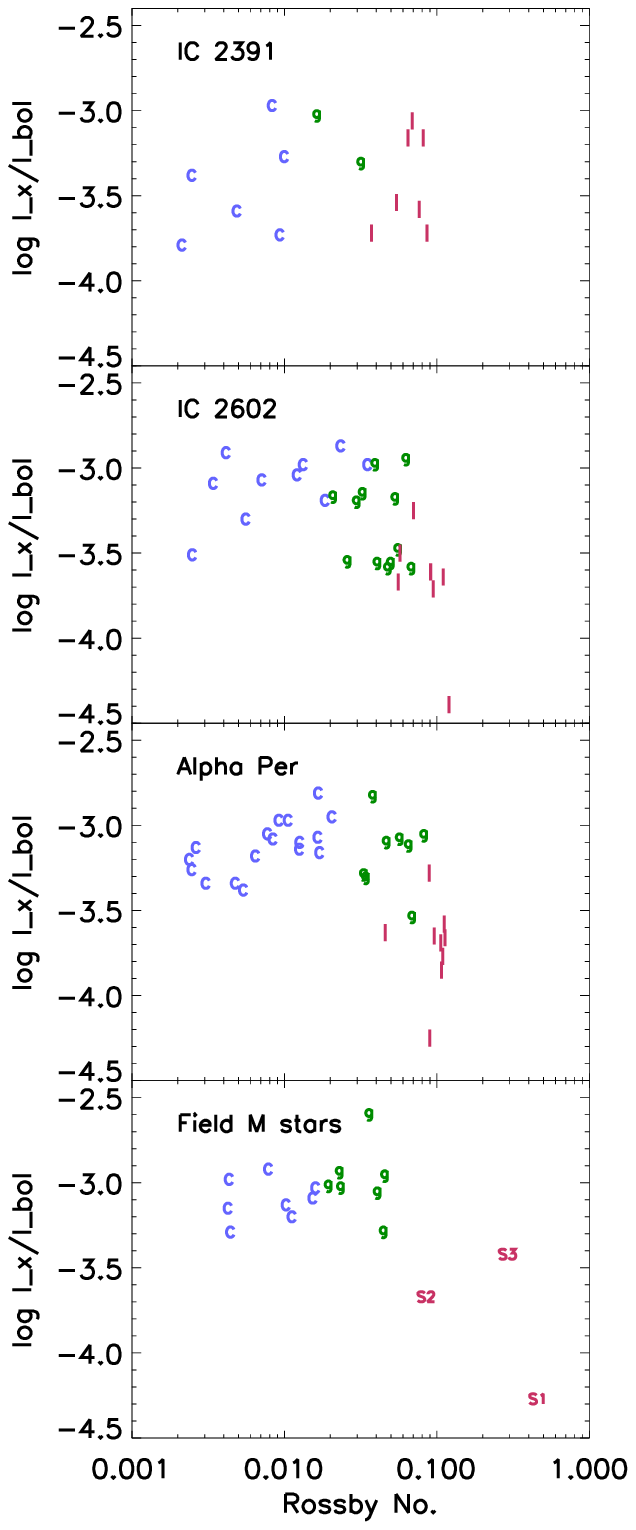}
\caption{Left hand panels: Color-period diagrams for the three open clusters 
IC\,2391, IC\,2602, Alpha Per, and the field M\,stars of James et al. (2000). 
Right hand panels: X-ray emission from the same stars plotted against Rossby 
number. Stars are marked C, g, or I, following the classification suggested 
previously in Barnes (2003). It appears that the three-segmented  X-ray 
morphology mirrors the classification of rotating stars into three groups. 
\label{fig1a}}
\end{figure}









\begin{thebibliography}{}
\bibitem[Barnes (2003)]{b03} Barnes, S.A., 2003, ApJ, 586, 464
\bibitem[Barnes et al. (1999)]{bsps99} Barnes, S.A., Sofia, S., Prosser, C.F. 
	\& Stauffer, J.R., 1999, ApJ, 516, 263
\bibitem[Donahue (1998)]{d98} Donahue, R.A., 1998, in ASP Conf. Ser. 154, 
	Tenth Cambridge Workshop on Cool Stars, Stellar Systems and the Sun, 
	ed. R.A. Donahue and J.A. Bookbinder, 1235
\bibitem[Feigelson et al. (2003)]{fei03} Feigelson, E.D., Gaffney III, J.A., 
	Garmire, G., Hillenbrand, L.A. \& Townsley, L., 2003, ApJ, in press
	(astro-ph/0211049)
\bibitem[James et al. (2000)]{jam00} James, D.J., Jardine, M.M., Jeffries, R.D.,
	Randich, S., Cameron, A.C. \& Ferreira, M., 2000, MNRAS, 318, 1217
\bibitem[Kim \& Demarque (1996)]{kd96} Kim, Y.-C. \& Demarque, P.D., 1996, 
	ApJ, 457, 340
\bibitem[Patten \& Simon (1996)]{ps96} Patten, B.M. \& Simon, T., 1996, ApJS,
	106, 489
\bibitem[Prosser \& Grankin (1997)]{pg97} Prosser, C.F. \& Grankin, K., 1997,
	CfA preprint 4539
\bibitem[Prosser et al. (1996)]{pea96} Prosser, C.F., Randich, S., Stauffer, 
	J.R., Schmitt, J.H.M.M. \& Simon, T., 1996, AJ, 112, 1570
\bibitem[Randich (2000)] {r00} Randich, S., 2000, in ASP Conf. Ser., 198, 
	Stellar Clusters and Associations: Convection, Rotation, and Dynamos, 
	ed. R. Pallavicini, G. Micela \& S. Sciortino (San Francisco: ASP), 401
\bibitem[Randich et al. (1995)] {rea95} Randich, S., Schmitt, J.H.M.M., 
	Prosser, C.F., \& Stauffer, J.R., 1995, A\&A, 300, 134
\bibitem[Randich et al. (1996)] {rea96} Randich, S., Schmitt, J.H.M.M., 
	Prosser, C.F., \& Stauffer, J.R., 1996, A\&A, 305, 785
\bibitem[Skumanich (1972)]{sku72} Skumanich, A., 1972, ApJ, 171, 565
\bibitem[Soderblom et al. (1991)]{sdj91} Soderblom, D.R., Duncan, D.K., \&
	Johnson, D.R.H., 1991, ApJ, 375, 722
\bibitem[Wilson (1963)]{w63} Wilson, O.C., 1963, ApJ, 138, 832

\end{thebibliography}
\end{document}